# Depth concentrations of deuterium ions implanted into some pure metals and alloys

*A.Yu. Didyk[1], R. Wiśniewski[2], K.Kitowski[2], V.Kulikauskas[3], T.Wilczynska[2], A.A. Shiryaev[4], Ya.V. Zubavichus[5]*

[1] G.N. Flerov Laboratory of Nuclear Reactions, Joint Institute for Nuclear Research, Joliot-Curie str., 6, 141980 Dubna, Russia, didyk@jinr.ru
[2] Institute of Atomic Energy-POLATOM, 05-400 Otwock/Świerk,
roland.wiśniewski@gmail.com; teresa.wilczynska@gmail.com, kacper.co@gmail.com
[3] D.V. Skobeltsin Institute of Nuclear Physics of Moscow State University, Lenisky Gory, 1, 119992 Moscow, Russia, vaclav@anna19.sinp.msu.ru
[4] Institute of Physical Chemistry and Electrochemistry RAS, Leninsky pr. 31, 119991 Moscow, Russia, a_shiryaev@mail.ru
[5] Russian Research Center "Kurchatov Institute", Kurchatov sq. 1, Moscow, Russia

**Abstract.** Pure metals (Cu, Ti, Zr, V, Pd) and diluted Pd-alloys (Pd-Ag, Pd-Pt, Pd-Ru, Pd-Rh) were implanted by 25 keV deuterium ions at fluences in the range $(1.2 \div 2.3) \times 10^{22}$ $D^+/m^2$. The post-treatment depth distributions of deuterium ions were measured 10 days and three months after the implantation using Elastic Recoil Detection Analysis (ERDA) and Rutherford Backscattering (RBS). Comparison of the obtained results allowed to make conclusions about relative stability of deuterium and hydrogen gases in pure metals and diluted Pd alloys. Very high diffusion rates of implanted deuterium ions from V and Pd pure metals and Pd alloys were observed. Small-angle X-ray scattering revealed formation of nanosized defects in implanted corundum and titanium.

## 1. Introduction

Behaviour of hydrogen isotopes in solids is of considerable fundamental and applied interest. Hydrogen-based energetics is a viable alternative to modern hydrocarbon-based technologies. One of the promising approaches involves hydrogen stored in metal hydrides [1-4]. Hydrogen and its heavier isotopes serve as a nuclear fuel in Fusion Reactor Power Plants [5-9]. They are also widely used in modern nuclear reactors for slowing down of neutrons, and as neutrons reflectors-mirrors, as safety materials and in the regulatory systems [7]. Production of high neutron fluxes for various applications remains an important challenge. The increase of capacity of metal foils to retain deuterium in the near-surface layers is another important problem. The basic challenge in all these applications is to achieve the highest possible hydrogen concentration in the employed materials preserving, in the same time, the capability to desorb hydrogen reversibly. Materials with high deuterium concentrations are important for some nuclear applications, such as neutron sources.

Ion implantation is one of the possible ways to create high hydrogen concentrations in the subsurface layers. In this work we report some results about behaviour of deuterium in various pure metals and alloys. We have studied in detail depth distribution of implanted deuterium; temporal stability of the implanted layers (for review of previous studies see [8-13] and references therein). We have monitored evolution hydrogen and deuterium concentrations after long time saturation of pure palladium and its diluted alloys. Finally the maximal stable concentrations of implanted deuterium ions in the metals and chemical compounds were estimated and compared with maximum equilibrium relations between metals and hydrogen (for example comparison with metal hydrides such as $TiH_2$, $MgH_2$, $AlH_3$ [7]).

## 2. Experimental methods and sample preparation

All samples of metals (purity of about 99.95 %) were prepared by mechanical polishing and then by electrochemical etching to achieve high quality smooth surfaces. Surface structures were studied using optical microscopy (metallographic optical microscope "Neophot", MOM), scanning electron microscopy (JSM-840, SEM) and atomic force microscopy (AFM) prior to the implantation to check the homogeneity.

Single crystal ingots of Pd and Pd alloys were grown at Institute of Atomic Energy (Świerk, Poland). The ingots were cut to thickness about 1÷2 mm. Then the samples were cold rolled, and finally mechanically and electrochemically polished to the final thickness of 50 ÷ 150 μm. For every compound four specimens were prepared. The first set was saturated by deuterium in high pressure chamber [14, 15]; the second part was implanted by $D^+$ ions up to high fluences. The third and fourth samples were annealed at temperature about 900K for 1 hour and then were deuterium saturated (the third specimen) and implanted up to the same ion fluences (the fourth).

## 2.1. Implantation of 25 keV $D^+$ ions at high fluences

The implantation of 25 keV deuterium ions was carried out at dedicated low energy ion irradiation beam line equipped with a separation magnet, deviation systems in both vertical and horizontal directions and a number of flux measuring Faraday cups in both directions [16].

Homogeneity of irradiation in a square of about 6×4 cm$^2$ was better than ±5%. All samples were glued by a heat conducting glue to a cooling substrate and temperature of implantation was **T** ≈30$^0$C. It is necessary to note that the ion beam flux during the implantation was about 3.5×10$^{17}$ D$^+$/(m$^2$×c), and the beam power was less than **W**≈0.075 W/cm$^2$. Maximum implantation fluence in all targets was **Φ$_{max}$** = 2.3×10$^{22}$ D$^+$/m$^2$ to exclude surface phenomena such as blistering and exfoliation as observed by optical and electron microscopies. Duration of the implantation for this maximum **D$^+$** fluence was about 20 hours. All metal samples and alloys were irradiated up to four ion fluences, i.e.: **Φ$_1$**=1.2×10$^{22}$, **Φ$_2$**=1.5×10$^{22}$, **Φ$_3$**=1.8×10$^{22}$ and **Φ$_{max}$**=2.3×10$^{22}$ D$^+$/m$^2$. Atomic force microscopy (AFM) was used for estimations of surface sputtering rates of all samples. The post-experimental metal surface roughness was comparable to calculations of the surface sputtering with computer code SRIM-2008 [17]. It is necessary to note that all implanted metal-samples were deformed by a spontaneous emergence of gas pores or gas bubbles (so called swelling processes). There was a surface discoloration with intensity depending on the ion implantation fluence.

## 2.2. Depth distribution of implanted D$^+$ ions in various materials: Elastic Recoil Detection Analysis (ERDA-RBS)

The depth distribution of implanted D$^+$ ions were measured using RBS setup based on electrostatic generator (EG-5) of the I.M. Frank Laboratory of Neutron Physics (FLNP, JINR, Dubna) [19] and tandetron at Skobeltsin Institute of Nuclear Physics (SINP, MSU, Moscow) [21]. Concentration profiles of light H/D atoms were measured by elastic recoil analysis (ERDA). The depth distributions of chemical complex compounds, such as diluted Pd alloys, Al$_2$O$_3$ and stainless steel (Fe$_{72}$Cr$_{18}$Ni$_{10}$ - SS) were measured by RBS [19-21]. We have employed 2.3 MeV He$^+$, the incidence angle was Θ=15°. Registration of scattering hydrogen and deuterium recoils was performed at 30° to an initial He$^+$ beam direction. The beamspot on the sample was elliptical with axes of a$_1$≈1 mm and a$_2$ 3.9 mm. All experimental spectra were then fitted with SIMNRA 6.05. The experimental ERDA and simulated spectra of H and D recoils under analyzing by He$^+$ ions with the energy of 1.9 MeV for Pd$_{0.9}$Pt$_{0.1}$ alloy implanted by deuterium ions with an energy of 25 keV at fluence of **Φ$_1$**=1.2×10$^{22}$ D$^+$/m$^2$ versus the number of energetic channels is presented in Fig.1. Here the sliding angle to surface is $\Theta=15°$ and angle of recoil registration to direction of $He^+$ ions is $\Omega=30°$. The thickness of $He^+$ ion adsorption mylar foil was 12 μm.

2.3 X-ray investigations of the samples.

The X-ray diffraction patterns were acquired from the Pd alloys foils in reflection geometry using X'Pert Panalytical powder diffractometer. Not surprisingly, the thickness of the implanted layers was too small to be detected. The main result of the diffraction experiments is the observation of texture, imposed by cold rolling of the samples.

Small angle X-ray scattering (SAXS) patterns were acquired at the STM beamline [27] at the Kurchatov synchrotron source using linear position-sensitive detector. Relatively high energy of incident monochromatic radiation (15-17 keV) was employed to permit recording of the patterns in transmission geometry. The samples were in vacuum. The size distribution of the scatterers was calculated in the assumption of spherical shape using GNOM software [28]. Comparison with unimplanted samples showed that the texture of the samples was not important for the scattering and the analysed signal is due to defects due to implanted/infused hydrogen.

3. Experimental results
3.1 **Depth distribution of implanted $D^+$ in some pure metals and some alloys**

Table 1 presents the most important calculated parameters for $D^+$ (25 keV) and for analyzing $He^+$ (2.3MeV) ions. Namely, mass and atomic densities, projected ranges of $D^+$ and $He^+$ ions and sputtering coefficients of all metals and chemical compounds studied are given. The concentration profile of implanted ions into a target can be calculated using the expression (see [21, 22]):

$$C(Z) = \frac{N_0}{2 \times S} \left[ erf\left( \frac{Z + \frac{\Phi \times S}{N_0} - R_p}{\sqrt{2} \times \Delta R_p} \right) - erf\left( \frac{Z - R_p}{\sqrt{2} \times \Delta R_p} \right) \right],$$

(1)

here $N_0$ - atomic target density (at/cm$^3$), $\Phi$ - ion fluence (ions/cm$^2$), S - sputtering coefficient for each target atoms, $R_p^D$ and $\Delta R_p$ are projected range and struggling of ions in target. Total implantation dose can be calculated using expression (1) as:

$$C_D = \int_0^\infty C(Z)dZ \approx \sum_{i=1}^{i=i_{max}} C_i \times \Delta Z_i$$

(2)

and maximum ions concentration should be

$$C^{max} = \frac{N_0}{S} \times erf\left[ \frac{\Phi \cdot S}{2\sqrt{2} \cdot N_0 \cdot \Delta R_p} \right]$$

(3)

at the depth $Z_{max} = R_p - \frac{\Phi \cdot S}{2 \cdot N_0}$.

### *Implantation to Zr target*

The depth-dependent concentrations of D and H (intrinsic impurity) at two implantation fluences: **$\Phi_2$**=1.5×10$^{22}$ D$^+$/m$^2$ and **$\Phi_{max}$**=2.3×10$^{22}$ D$^+$/m$^2$ for Zr samples are presented in Fig. 2. Integral concentrations of the implanted ions can be introduced using simple expressions $c_{D,H} \equiv \sum_{i=1}^{i_{max}} c_i(D,H) \times \Delta Z_i$, where indexes "$i$" correspond to depth layers in a target from Z=0 up to maximum ERDA measured depth $Z < R_p^{He}$. As is seen from Table 1, the projected range of D ions in Zr foil is equal to $R_p^D = 1923 \pm 643$ Å. Using sputtering coefficient $S_{Zr} = 2.48 \times 10^{-3} \frac{at.Zr}{at.D^+}$ it is very easy to estimate the thickness of the surface layer sputtered by implantation of D$^+$ at a fluence of $\Phi_{max}$=2.3×10$^{22}$ D$^+$/m$^2$ to be equal to $\Delta Z = \Phi_{max} \times S_{Zr} / N_{Zr} = 13.3$ Å.

Figure 2 shows that a) the depth concentrations of D atoms for all fluences have very large spread along the implanted ion path and the width of implanted zone is much larger then the projected range $R_p^D = 1923 \pm 643$ Å; b) the total concentrations of D atoms correspond closely to the experimental ion fluences for Zr targets only; c) the maximum obtained concentration of D atoms in Zr is about 47%, i.e. $n_{Zr}^{exper}(D^+) = 0.47 \frac{at.D}{at.Zr} < n_{Zr}^{max} = 2.95 \frac{at.D}{at.Zr}$. Here $n_{Zr}^{max}(D^+) = 2.95 \frac{at.D}{at.Zr}$ is the calculated deuterium-metal ratio for maximum fluence of implantation, i.e. $n^{max} \equiv C^{max}/N_0$. The calculated profiles and another parameters were obtained using SRIM-2008 simulation [17] (depth profiles, $R_p$, $\Delta R_p$ and $N_0$) and using homemade computer program SPECTR based on expressions (1)-(3).

### *Implantation to Ti target*

The measured depth distributions of D and H atoms in implanted Ti samples (ion fluences $\Phi_3$=1.8×10$^{22}$ D$^+$/m$^2$ and $\Phi_{max}$=2.3×10$^{22}$ D$^+$/m$^2$) are presented in Fig.3. *In contrast to implantation to Zr metal the integral concentration of implanted deuterium is lower due to larger projected range (see Table 1). This could be the reason for wider spread of the profiles.*

### *Implantation to Al$_2$O$_3$ target*

The total concentration of implanted deuterium into synthetic Al$_2$O$_3$ single crystal is shown in Fig.4. These crystals possessed high surface quality as shown by AFM. The use of ERDA technique with relatively low projected range of He$^+$ (2.3 MeV) ions and small thickness of implanted layers required a smooth surface. It is easy to see from Fig.4 that: a) the depth distributions of implanted D$^+$ ions for fluence $\Phi_{max}$=2.3×10$^{22}$ D$^+$/m$^2$ have a very big spread along the implanted ion path and the width of implanted zone is much larger than the projected range $R_p^D = 2832 \pm 602$ Å; b) the integral concentration of D atoms exceeds $C_D$>6.43×10$^{21}$ D$^+$/m$^2$;

c) maximum measured concentration of D atoms in $Al_2O_3$ sample is about 15%, i.e. $n_{Al_2O_3}^{exper}(D^+) = 0.15 \frac{at.D}{at.Al_2O_3} \ll n_{Al_2O_3}^{max} = 1.69 \frac{at.D}{at.Al_2O_3}$. Here $n_{Al_2O_3}^{max} = 1.69 \frac{at.D}{at.Al_2O_3}$ is the calculated deuterium/$Al_2O_3$ ratio for the maximal implantation fluence. The maximum concentrations of D atoms were achieved at the depths between 0.26 and 0.4 μm. This depth region is comparable with the projected range of $D^+$ ions in $Al_2O_3$ sample $R_p^D = 0.28\pm0.06$ μm.

The experimental low value of the deuterium/$Al_2O_3$ ratio can be explained by a high spread of the implantation layer, larger than the analyzed $He^+$ ion projected range. It is necessary to check this conclusion by another method sensitive to deuterium depth distribution, e.g. by secondary mass ion spectrometry.

***Implantation to* copper and stainless steel *targets***

Figure 5 shows the depth profiles of D and H atoms in copper and stainless steel ($Cr_{18}Ni_{10}Fe_{72}$) samples at the $D^+$ fluence $\Phi_2=1.5\times10^{22}$ $D^+/m^2$ as measured by ERDA. One can conclude that total D concentration and shape of the depth profiles in these samples are similar to the titanium one (see above).

***Implantation to* vanadium *targets***

Very interesting results were obtained by the ERDA studies of vanadium foils implanted by $D^+$ ions for all studied fluences. The total concentrations of D atoms in implanted pure V and Pd and $Pd_{0.9}Rh_{0.1}$ alloy samples are presented in Fig.6 and Fig.7 for comparison.

The irradiated surfaces of V and Pd samples implanted by $D^+$ ions at a fluence $\Phi_{max}=2.3\times10^{22}$ $D^+/m^2$ became dark brown and dark blue, respectively.

The calculated D/V and D/Pd ratios for fluences of implantation $\Phi_{max}=2.3\times10^{22}$ $D^+/m^2$ and $\Phi_1=1.2\times10^{22}$ $D^+/m^2$ (3) have values $n_V^{max}(D^+) = 2.06 \frac{at.D}{at.V}$ and $n_{Pd}^{max}(D^+) = 1.69 \frac{at.D}{at.Pd}$, respectively. The total concentrations and depth concentrations of implanted $D^+$ ions are very low. The measured values of D concentrations in both V and Pd implanted foils are less than 1÷2 % only, i.e. these values are near the detection limit of the employed ERDA setup.

It is known that Pd foils can be used as superfilters of hydrogen isotopes and impermeable membranes for other gases [7, 14, 15, 18, 20, 24-26]. The conclusion of our studies is that V foils can be also used as selective membranes for purification of hydrogen isotopes from other gases. It was shown [26] that with the use of so-called atomizer, i.e. special heating element (1000 ÷2300 K temperature range) which allows one to case dissociation of molecular hydrogen and its heavier isotopes (chemical reaction $H_2/D_2/T_2 \rightarrow H+H/D+D/T+T$) the membranes made of vanadium together with niobium can also purify hydrogen. Experiments on vanadium-assisted gas separation began only recently. To our knowledge no publications are devoted to investigation of desorption of high amounts of hydrogen-deuterium atoms after high dose implantation.

All experimentally measured values for D atom total concentrations in the studied materials such as Zr, Ti, V, Cu, Pd, SS and $Al_2O_3$ are summarised in Table 2. It is important to note the sputtered layers in all studied metals and alloys are very thin (about 10÷20 Å) in comparison to the projected range of $D^+$ ions. As it can be seen in Table 2, there is a close relation between experimentally measured total concentrations and implanted ion fluences, though a large spread of experimental profiles is observed in metals and alloys targets. Some positions in Table 2 are not filled yet and there are some inconsistencies to be settled in the nearest future.

Capability of pure Pd to dissolve significant amounts of hydrogen was discovered long time ago (1888, see [7, 15]). Since that time detailed investigations of hydrogen behaviour in Pd alloys (in particular with noble metals) attracts considerable applied and fundamental interest. The D and H depth profiles in several Pd alloys were measured by ERDA repeatedly: the first measurement after 10 days and 3 months after the implantation (fig. 8) (see also fig. 7 for depth profiles for Pd and $Pd_{0.9}Rh_{0.1}$ alloy). From Fig. 8 one can conclude that the concentrations of implanted D atoms remain virtually unchanged. Some differences between the measured values can be connected also with different energies of ERDA setups at FLNP-JINR and at SINP-MSU: 2.3 MeV and 1.9 MeV, respectively. It is necessary to note that the observations of high H concentration at great depth should be verified. The retention of high total concentrations of D atoms in measured Pd and Pd alloys can be explained by absorbtion of D atoms in gas bubbles and at grain boundaries [9-13].

Table 3 summarises the ERDA results for total D and H concentrations in Pd and in some palladium based alloys as: $Pd_{0.9}Ag_{0.1}$, $Pd_{0.9}Pt_{0.1}$, $Pd_{0.9}Ru_{0.1}$ and $Pd_{0.9}Rh_{0.1}$. We repeat here some main results: from comparison of tables 1 and 2 it is clear that the concentrations of implanted deuterium and of intrinsic hydrogen do not change much for a long period (three months) between the measurements. Minor differences are most probably related by different experimental setups for ERDA study and different energies of employed $He^+$ ions beams at FLNP (JINR, Dubna) and at SINP (MSU, Moscow). It is necessary to note that the depth distributions of intrinsic hydrogen atoms are very deep and have high concentrations for all Pd-alloys. We conclude that both isotopes - D and H atoms – exist in alloy lattices as substitutional and interstitial atoms and/or as small gas bubbles [9-13]. The obtained hydrogen concentrations are high: $(1.5 \div 3.2) \times 10^{21}$ $H/m^2$ which is very promising for possible future applications. The depth distribution of intrinsic H atoms in all studied Pd-alloys was very broad.

## 3.2. Small-angle X-ray scattering

The small-angle scattering intensity for almost all studied foils was negligible which indicates high homogeneity of the electron density distribution in the samples. This can be easily interpreted as absence of extended defects and/or of clustering of point defects. However, in the corundum and titanium samples the SAXS signal was observed The size distribution curves of spherical scatterers for these samples is shown in Fig. 9. Scattering from initial (unimplanted) samples was used as a background, thus permitting to minimize any possible contribution form intrinsic defects and/or from texture.

The scattering patterns for these two samples are very different. The scatterers in $Al_2O_3$ produced by implantation of 25 keV $D^+$ ions up to dose of $\mathbf{\Phi_{max}}=2.3\times10^{22}$ $D^+/m^2$ can be described as a monodisperse system with diameters close to 1 nm. Such narrow size distribution is not very common for inorganic systems. Most likely, this scattering is due to small gas bubbles or similar defects. However, in titanium sample the size distribution of the scatterers is very different: it is broad with a main peak around 10 nm (radius) and secondary one around 16 nm. Unique assignment of these peaks using only SAXS data is impossible. The scattering could be assigned to clusters of point, e.g., radiation, defects. They might be precursors of bubbles, or reflect formation of hydride phase.

The SAXS data might be explained by very different behaviour of hydrogen in metals and ceramics. In metals hydrogen is present as a "lattice gas" or form hydrides. In ceramics hydrogen may form point defects saturating dangling bonds, it can recombine to $H_2$ molecule etc. In addition, the diffusion rates of H isotopes are much higher in metals, thus explaining high desorption rates.

It is well known [9-12, 22, 23] that deuterium and helium confined in small bubbles in metals with high mechanical strength can be present in solid state. The internal pressure should be about $P=2\times\gamma/R$, where $\gamma$ is a surface tension. The estimated values of He pressure in bubbles in aluminum and nickel is $P \approx 13$ MPa (radius of the bubble is R=0.65 nm with atomic density $\approx 1.4\times10^{23}$ He/cm$^3$) and $P \approx 50$ GPa (with atomic density $\approx 1.4\times10^{23}$ He/cm$^3$), respectively [12, 13, 22, 23].

## 4. Conclusions

The depth distributions of D and H atoms after implantation of $D^+$ into several pure metals (Zr, Ti, Cu, V and Pd), alloys (SS, $Pd_{0.9}Ag_{0.1}$, $Pd_{0.9}Pt_{0.1}$, $Pd_{0.9}Ru_{0.1}$ and $Pd_{0.9}Rh_{0.1}$) and single crystal $Al_2O_3$ were measured using ERDA method.

It is shown that total concentrations of implanted deuterium for high fluences $\Phi=1.2\times10^{22}$, $1.5\times10^{22}$, $1.8\times10^{22}$ and $2.3\times10^{22}$ $D^+/m^2$ correspond to employed ion fluences, in particular for Zr foils. The observed depth profiles of deuterium shows a spread along the projected range incident ions.

$D^+$ implantation up to very high fluences with small ion flux $\approx 3.5\times10^{17}$ $D^+/(m^2\times sec)$ allowed to obtain high total concentrations in most studied metals and alloys except for V and Pd. Deuterium-saturated layers span considerable depth, several times the projected implantation ranges. No blistering and exfoliation was observed.

High desorption losses of deuterium were observed for implanted vanadium samples for all used fluences. The comparison of implanted samples of V and Pd allowed to conclude that V foils can be used together with more expensive Pd foils for separation and purification of hydrogen and its heavier isotopes from other gases.

In future it is important to study deuterium-implanted other elements of V group of Periodic Table as Nb and Ta. Vanadium is known to have lower level and shorter period of fission fragment radioactivity decays than heavier elements of V-th group, so this metal is much better suited for applications in fusion reactors.

Desorption of implanted deuterium as well as its spread are very small for Pd and its diluted alloys. These samples also contain high concentrations of intrinsic hydrogen.

The metal surfaces change color once a certain critical fluence value is surpassed. The extent of the changes are proportional to implantation fluence. This allows to estimate the implantation fluence at least seni-quantitatively, e.g. using spectrophotometer.

All implanted metal samples were deformed, probably by emergence of gas pores or gas bubbles.

Small angle X-ray scattering (SAXS) reveals appearance of small well-defined heterogeneities with radii approx. 1 nm in implanted corundum. These defects might correspond to gaz bubbles. Heterogeneities also appear in implanted titanium, though the size distribution is much broader and the defects are much larger (>10 nm). These scatterers might represent early stages of bubbles formation or agglomeration of point defects. No SAXS signal was detected for other implanted samples, indicating absence of defects clusterisation.

**Acknowledgements**

Authors are thankful to Prof. Andrzej Czachor (Atomic Energy Institute - POLATOM, Šwerk/Otwock, Poland) for the provision of the Pd and Pd based alloy materials for the studies and fruitful discussions in all stages of studies, and to Prof. Stanislaw Filipek (Institute of Chemical Physics, Warsaw, Poland) for provision of high pressure set-up for metals and alloys saturations.

Figure captions 1-9:

Fig.1. The experimental ERDA and simulated spectra of H and D recoils in $Pd_{0.90}Pt_{0.10}$ alloy implanted by 25 keV deuterium ions up to fluence of $\mathbf{\Phi_1}$=1.2×10$^{22}$ D$^+$/m$^2$.

Fig.2. The depth profiles of D and H atoms in Zr samples after 25 keV D$^+$ implantation at two fluences: $\mathbf{\Phi_2}$=1.5×10$^{22}$ D$^+$/m$^2$ (**a**) and $\mathbf{\Phi_{max}}$=2.3×10$^{22}$ D$^+$/m$^2$ (**b**).

Fig.3. The depth profiles of D and H atoms in Ti samples after 25 keV $D^+$ implantation at two fluences: $\mathbf{\Phi_2}$=1.8×10$^{22}$ ion/m$^2$ (**a**) and $\mathbf{\Phi_{max}}$=2.3×10$^{22}$ ion/m$^2$ (**b**).

Fig.4. The depth profiles of D and H atoms in $Al_2O_3$ single crystal after 25 keV $D^+$ implantation at fluence $\mathbf{\Phi_{max}}$=2.3×10$^{22}$ $D^+$/m$^2$ (**b**).

Fig.5. The depth profiles of D and H atoms after 25 keV $D^+$ implantation at one fluence: $\mathbf{\Phi_2}$=1.5×10$^{22}$ $D^+$/m$^2$ in Cu (**a**) and Stainless steel-$Cr_{18}Ni_{10}Fe_{72}$ (**b**) samples.

Fig.6. The depth profiles of D and H atoms in V samples after 25 keV $D^+$ implantation at two fluences: $\mathbf{\Phi_2}$=1.5×10$^{22}$ $D^+$/m$^2$ (**a**) and $\mathbf{\Phi_{max}}$=2.3×10$^{22}$ $D^+$/m$^2$ (**b**).

Fig.7. The depth profiles D and H atoms after 25 keV $D^+$ implantation at fluence $\mathbf{\Phi_1}$=1.2×10$^{22}$ $D^+$/m$^2$ in pure Pd (**a**) and $Pd_{0.9}Ag_{0.1}$ (**b**) samples.

Fig.8. The depth profiles of D and H atoms after 25 keV $D^+$ implantation at fluence $\mathbf{\Phi_1}$=1.2×10$^{22}$ $D^+$/m$^2$ in the following samples $Pd_{0.9}Ag_{0.1}$ (**a, b**); $Pd_{0.9}Pt_{0.1}$ (**c, d**); $Pd_{0.9}Rh_{0.1}$ (**e, f**); $Pd_{0.9}Ru_{0.1}$ (**g, h**). The left column – 10 days; the right column – 3 month after the implantation.

Fig. 9. Size distribution of scatterers in ion implanted titanium and corundum.

Captions to Tables 1 - 3:

Table 1. Calculated and SRIM-2008 parameters for all metals and some chemical compositions.

Table 2. Complete total concentrations of D atoms after implantation in some metals and alloys at various fluences.

Table 3. Complete total concentrations of D (top value) and H atoms (bottom value) and after implantation of $D^+$ ions for all studied palladium alloys: $Pd_{0.9}Ag_{0.1}$. $Pd_{0.9}Pt_{0.1}$. $Pd_{0.9}Ru_{0.1}$ and $Pd_{0.9}Rh_{0.1}$ at a fluence $\Phi_1=1.2\times10^{22}$ $D^+/m^2$.

Table 1. Calculated (SRIM-2008 [17]) parameters for all studied compounds.

| Material | Zr | Ti | Cu | V | Pd | SS | Al$_2$O$_3$ |
|---|---|---|---|---|---|---|---|
| Implantation of 25 keV deuterium ions | | | | | | | |
| $\rho$, g/cm$^3$ | 6.49 | 4.519 | 8.92 | 5.96 | 12.03 | 7.849 | 3.5 |
| N, at.M/cm$^3$, ×10$^{22}$ | 4.284 | 5.681 | 8.453 | 7.045 | 6.803 | 8.526 | 10.33 |
| $R_p^D$, Å | 1923±643 | 2444±695 | 1854±673 | 1955±571 | 1237±458 | 1998±698 | 1923±643 |
| S, at.M/at.D | 0.00248 | 0.0019 | 0.010 | 0.0025 | 0.0011 | 0.0012(Cr) 0.0007(Ni) 0.0049(Fe) | 0.00153(Al) 0.0024(O) |
| Analyzing 2.3 MeV helium ions with energy, grazing angle Θ=15° | | | | | | | |
| $R_p^D$, μm | 1.32 | 1.43 | 1.07±0.33 | 1.04 | 0.88±0.30 | 0.95±0.26 | 1.38 |

Table 2. Total measured concentrations of implanted deuterium in several metals and alloys at various fluences.

| Fluence of $D^+$ ion implantation, $\times 10^{22}$ $D^+/m^2$ | $\Phi_1=1.2$ | $\Phi_2=1.5$ | $\Phi_3=1.8$ | $\Phi_{max}=2.3$ |
|---|---|---|---|---|
| Zirconium, Zr | - | >1.5×10$^{22}$ $D^+/m^2$ No desorption! | ≈1.5×10$^{22}$ $D^+/m^2$ No desorption! | ~2.3×10$^{18}$ $D^+/cm^2$ No desorption! |
| Titanium, Ti | - | - | >0.6×10$^{22}$ $D^+/m^2$ Desorption? | >0.65×10$^{22}$ $D^+/m^2$ Desorption? |
| Copper, Cu | - | >2.9×10$^{21}$ $D^+/m^2$ Desorption (?) | - | - |
| Palladium, Pd | >4.5×10$^{20}$ $D^+/m^2$ Super H/D-filter | - | - | - |
| Vanadium, V | - | < 8.8×10$^{20}$ $D^+/m^2$ Super H/D-filter | ≈ 4.4×10$^{20}$ $D^+/m^2$ Super H/D-filter | ≈ 7.0×10$^{20}$ $D^+/m^2$ Super H/D-filter |
| Stainless steel, SS-Fe$_{72}$Cr$_{18}$Ni$_{10}$ | - | >2.3×10$^{21}$ $D^+/m^2$ Desorption (?) | - | - |
| Al$_2$O$_3$ | - | - | - | >6.4×10$^{22}$ $D^+/m^2$ Desorption? |

Table 3. Total concentrations of D (top value) and H atoms (bottom value) for all studied palladium alloys: $Pd_{0.9}Ag_{0.1}$. $Pd_{0.9}Pt_{0.1}$. $Pd_{0.9}Ru_{0.1}$ and $Pd_{0.9}Rh_{0.1}$ at fluence $\Phi_1=1.2\times10^{22}$ $D^+/m^2$.

| Total concentrations | Pure Pd | $Pd_{0,9}Ag_{0,1}$ | $Pd_{0,9}Pt_{0,1}$ | $Pd_{0,9}Ru_{0,1}$ | $Pd_{0,9}Rh_{0,1}$ |
|---|---|---|---|---|---|
| $\rho$, g/cm$^3$ | 12.02 | 11.87 | 12.96 | 12.05 | 12.03 |
| N, at.M/cm$^3$, ×10$^{22}$ | 6.803 | 6.798 | 6.770 | 6.854 | 6.803 |
| Implantation of deuterium ions, energy - 25 keV, fluence - $\Phi_1=1,2\times10^{22}$ $D^+/m^2$ | | | | | |
| $R_p^D$, Å | 1237±458 | 1267±470 | 1227±470 | 1255±466 | 1264±473 |
| S, at.M/at.D | 0.00112 | 0.0104(Pd) 0.00086(Ag) | 0.0083(Pd) 0.00062(Pt) | 0.0073(Pd) 0.00084(Ru) | 0.00074(Pd) 0.00098(Rh) |
| Equation (2) | n(D$^+$)≈1.42 at.D$^+$/at.M at depth ≈1450 Å | | | | |
| ERDA measurements after implantation (FLNP-JINR, He$^+$ -2,3 MeV) | | | | | |
| $R_p^{He}$, μm | 0.88±0.30 | 0.89±0.30 | 0.88±0.32 | 0.87±0.30 | 0.88±0.28 |
| $C_D$, D$^+$/m$^2$, ×10$^{21}$ $C_H$, H/m$^2$ ×10$^{21}$ | ~0.45 >2.41 | ~0.29 >1.70 | ~0.34 >1.57 | ~0.31 >2.39 | ~0.29 >3.21 |
| ERDA measurements two month later after implantation (MSU, He$^+$ -1,9 MeV) | | | | | |
| $C_D$, D$^+$/m$^2$, ×10$^{21}$ $C_H$, H/m$^2$ ×10$^{21}$ | - - | ~0.20 >1.10 | ~0.31 >1.04 | ~0.27 >0.71 | ~0.19 >1.10 |

FIGURE 1.

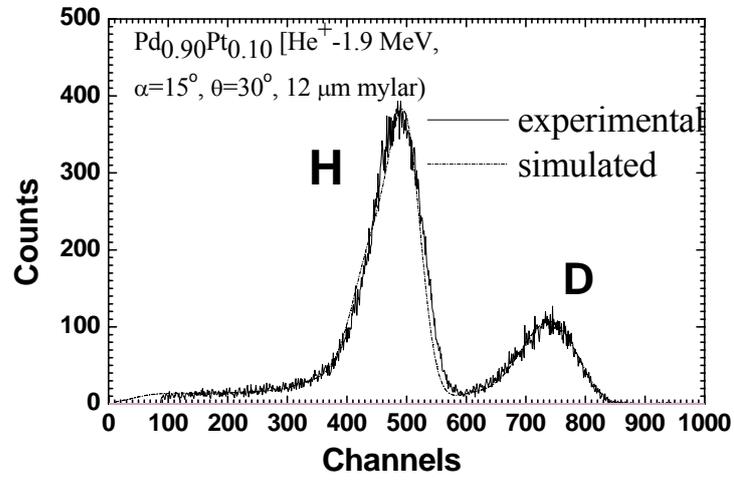

Fig.1. The experimental ERDA and simulated spectra of H and D recoils in $Pd_{0.90}Pt_{0.10}$ alloy implanted by 25 keV deuterium ions up to fluence of $\Phi_1 = 1.2 \times 10^{22}\,D^+/m^2$.

FIGURE 2.

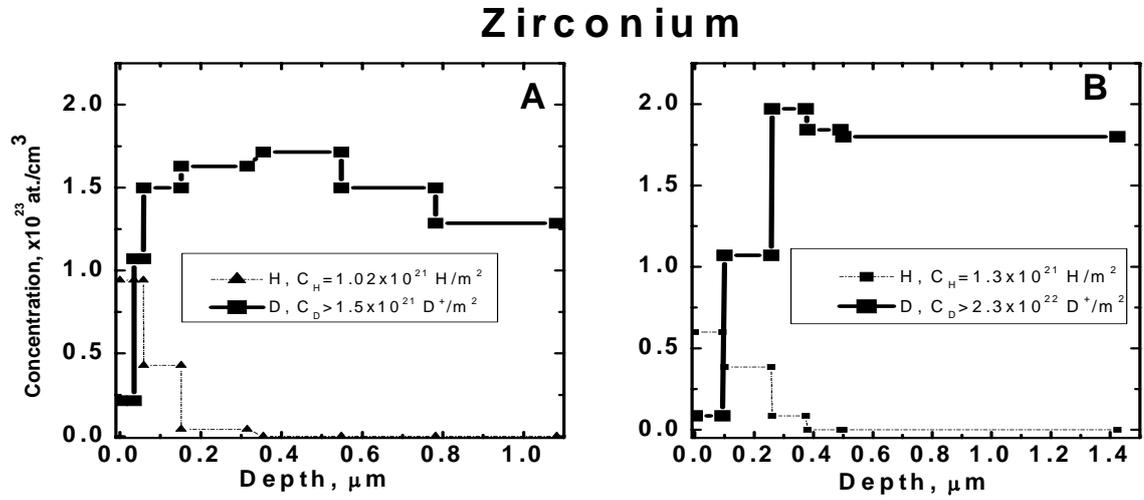

Fig.2. The depth profiles of D and H atoms in Zr samples after 25 keV $D^+$ implantation at two fluences: $\Phi_2=1.5\times10^{22}$ $D^+/m^2$ (**a**) and $\Phi_{max}=2.3\times10^{22}$ $D^+/m^2$ (**b**). Concentration of the isotopes are given. $D^+$ projected range is $R_p = 1923\pm643$ Å.

FIGURE 3.

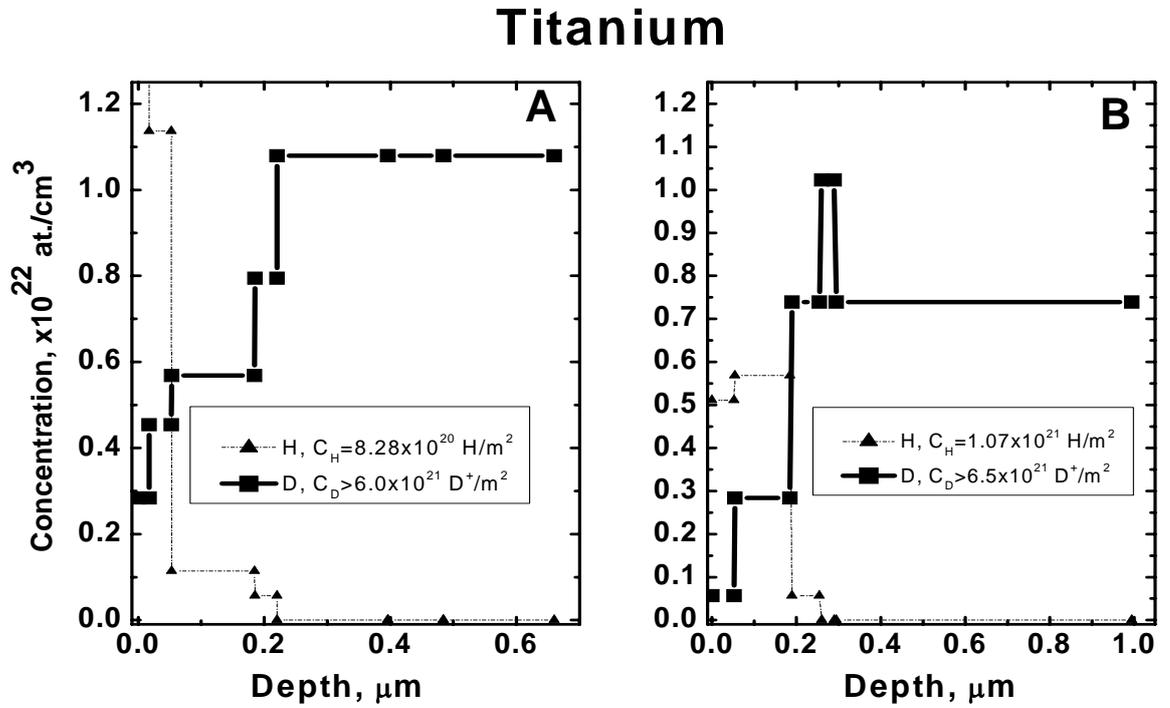

Fig.3. The depth profiles of D and H atoms in Ti samples after 25 keV $D^+$ implantation at two fluences: $\Phi_2=1.8\times10^{22}$ ion/m$^2$ (**a**) and $\Phi_{max}=2.3\times10^{22}$ ion/m$^2$ (**b**). $D^+$ projected range is $R_p = 2446\pm493$ Å.

FIGURE 4.

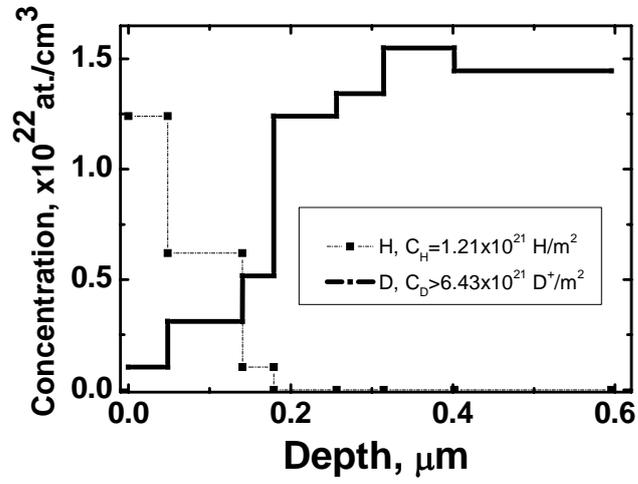

Fig.4. The depth profiles of D and H atoms in $Al_2O_3$ single crystal after 25 keV $D^+$ ion implantation at fluence $\Phi_{max}=2.3\times10^{22}\,D^+/m^2$ (**b**). $D^+$ projected range is $R_p = 2832\pm602$ Å

FIGURE 5.

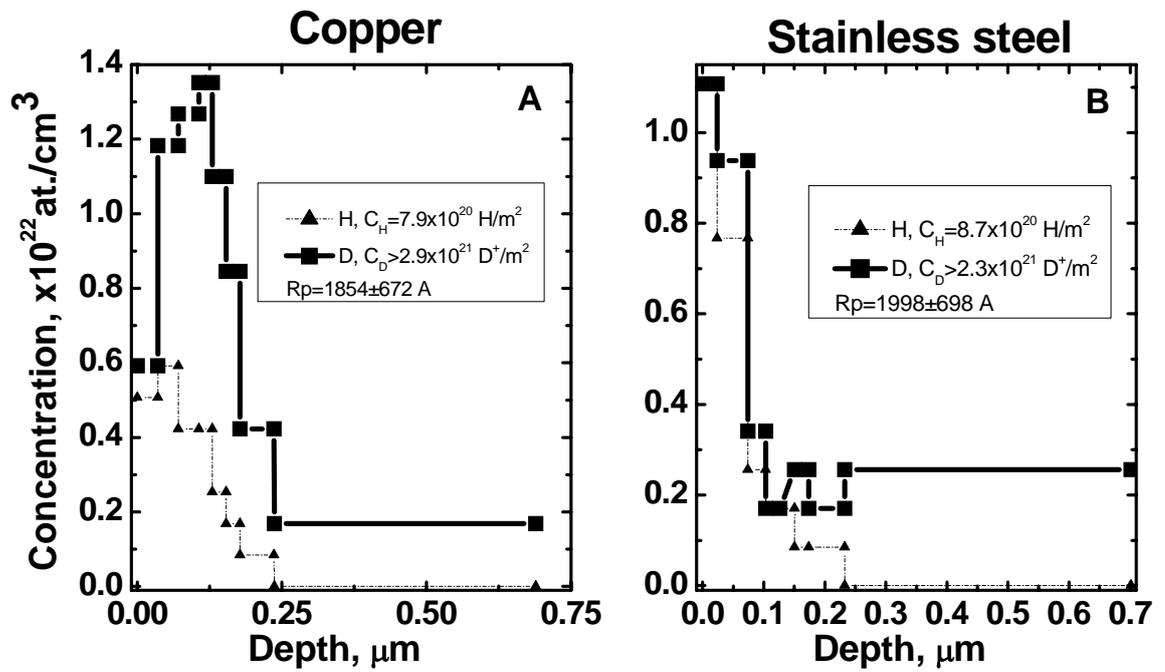

Fig.5. The depth profiles of D and H atoms after 25 keV $D^+$ ion implantation at one fluence: $\Phi_2$=1.5×10$^{22}$ $D^+/m^2$ in Cu (**a**) and Stainless steel-$Cr_{18}Ni_{10}Fe_{72}$ (**b**) samples.

FIGURE 6.

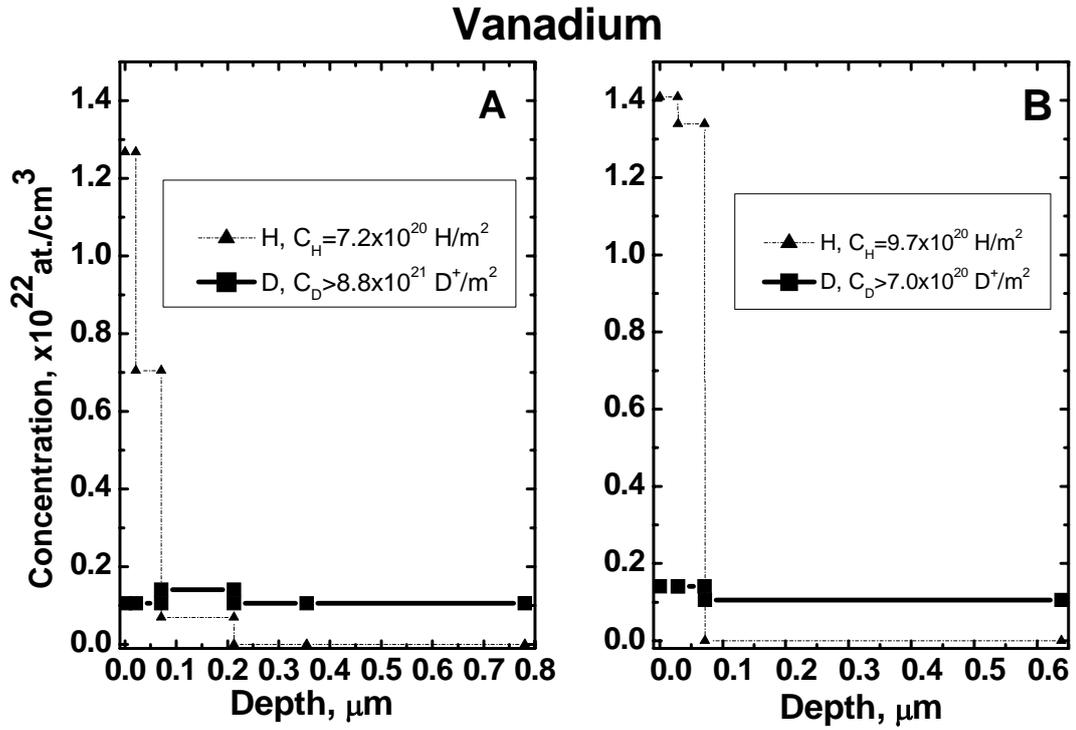

Fig.6. The depth concentrations of D and H atoms in V samples after 25 keV $D^+$ ion implantation at two fluences: $\Phi_2=1.5\times10^{22}$ $D^+/m^2$ (**a**) and $\Phi_{max}=2.3\times10^{22}$ $D^+/m^2$ (**b**). $D^+$ projected range is $R_p = 1955\pm571$ Å

FUGURE 7.

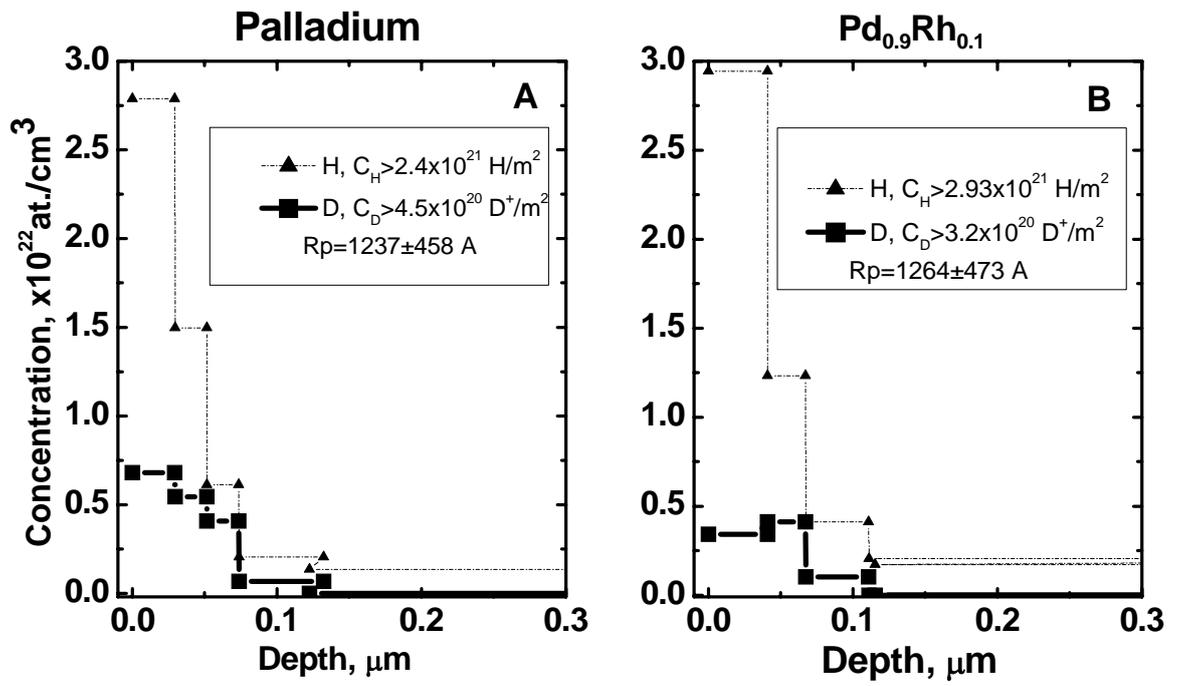

Fig.7. The depth concentrations of D and H atoms after 25 keV $D^+$ ion implantation at fluence $\Phi_1=1.2\times10^{22}$ $D^+/m^2$ in pure Pd (**a**) and $Pd_{0.9}Rh_{0.1}$ (**b**) samples.

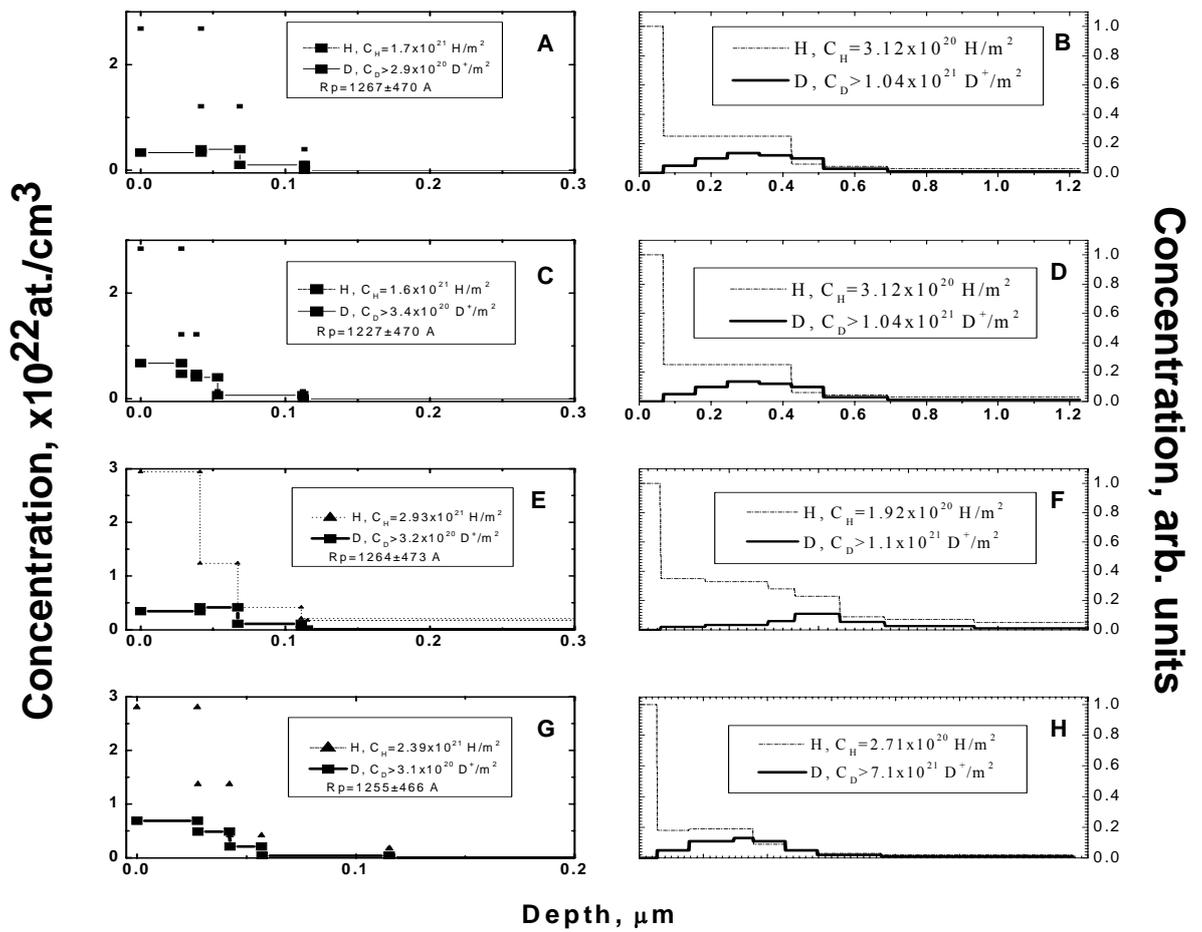

Fig.8. The depth profiles of D and H atoms after 25 keV $D^+$ implantation at fluence $\Phi_1=1.2\times10^{22}$ $D^+/m^2$ in the following samples $Pd_{0,9}Ag_{0,1}$ (**a, b**); $Pd_{0,9}Pt_{0,1}$ (**c, d**); $Pd_{0,9}Rh_{0,1}$ (**e, f**); $Pd_{0,9}Ru_{0,1}$ (**g, h**). The left column – 10 days; the right column – 3 month after the implantation.

Fig. 9. Size distribution of scatterers in ion implanted titanium and corundum.

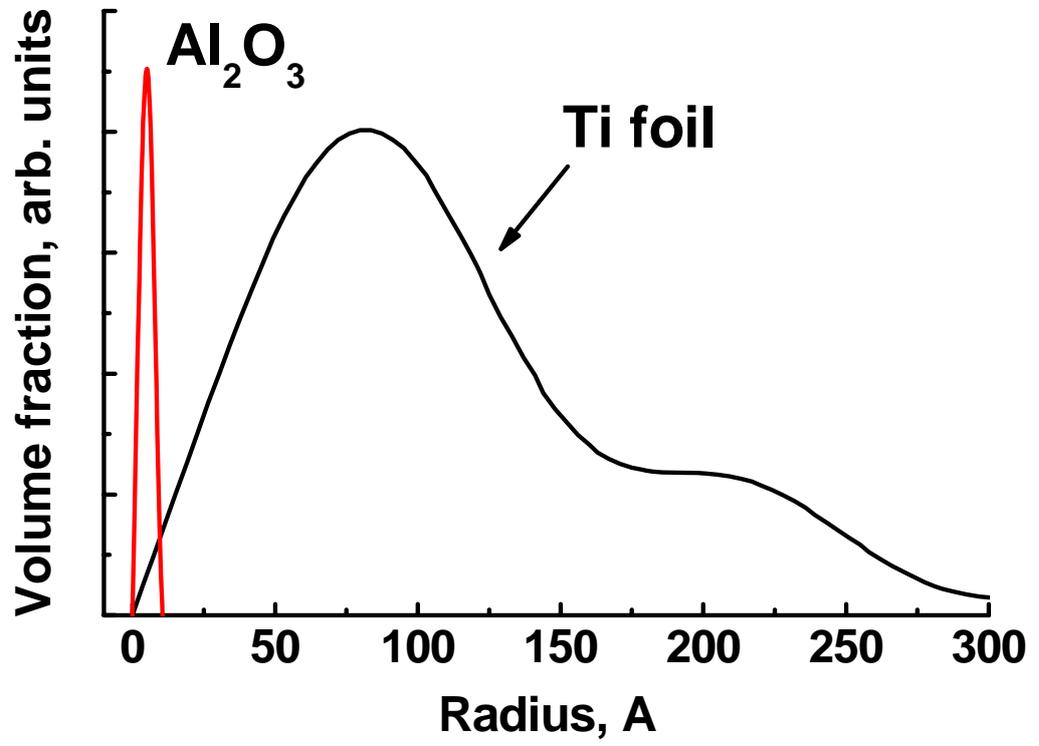